\documentclass[sigconf]{acmart}
\AtBeginDocument{%
  }

\setcopyright{acmlicensed}
\copyrightyear{2018}
\acmYear{2018}
\acmDOI{XXXXXXX.XXXXXXX}
\acmConference[Conference acronym 'XX]{Make sure to enter the correct
  conference title from your rights confirmation email}{June 03--05,
  2018}{Woodstock, NY}
\acmISBN{978-1-4503-XXXX-X/2018/06}

\usepackage{multirow}
\usepackage{algpseudocode}
\usepackage[ruled, vlined, linesnumbered]{algorithm2e}
\usepackage[utf8]{inputenc}
\begin{document}

\title{HoloRec: Holistic Encoding and Interleaved Reasoning  for Generative Recommendation}


\author{Shuqi Zhao}
\authornote{Both authors contributed equally to this research.}
\email{zhaoshuqi22@mails.ucas.ac.cn}
\affiliation{%
  \institution{Institute of Information Engineering, Chinese Academy of Sciences}
  \city{Beijing}
  \country{China}
}

\author{Jingsong Su}
\authornotemark[1]
\email{sujingsong@mail.bnu.edu.cn}
\affiliation{%
  \institution{School of Artificial Intelligence, Beijing Normal University}
  \city{Beijing}
  \country{China}
}
\author{Xiang Liu}
\email{liuxiang@iie.ac.cn}
\affiliation{%
  \institution{Institute of Information Engineering, Chinese Academy of Sciences}
  \city{Beijing}
  \country{China}
}
 
\author{Xingzhi Yao}
\author{Yiming Qiu}
\author{Huimu Wang}
\email{\{yaoxingzhi,qiuyiming\}@jd.com}
\email{\{wanghuimu1\}@jd.com}
\affiliation{%
  \institution{JD.com,Beijing,China}
  \city{}
  \country{}
} 
\author{Liang Lin}
\author{Pengbo Mo}

\email{\{linliang,mopengbo\}@iie.ac.cn}
\affiliation{%
  \institution{Institute of Information Engineering, Chinese Academy of Sciences}
  \city{Beijing}
  \country{China}
}

\author{Mingming Li}\authornote{Corresponding authors.}

\author{Jiao Dai}
\email{\{limingming,daijiao\}@iie.ac.cn}
\affiliation{%
  \institution{Institute of Information Engineering, Chinese Academy of Sciences}
  \city{Beijing}
  \country{China}
}

\author{Jizhong Han}
\email{hanjizhong@iie.ac.cn}
\affiliation{%
  \institution{Institute of Information Engineering, Chinese Academy of Sciences}
  \city{Beijing}
  \country{China}
}

\author{Songlin Hu}
\email{husonglin@iie.ac.cn}
\affiliation{%
  \institution{Institute of Information Engineering, Chinese Academy of Sciences}
   \city{Beijing}
  \country{China}
}
 \renewcommand{\shortauthors}{Shuqi Zhao et al.}

\begin{abstract}
Generative recommendation models that formulate the task as sequence generation overcome the objective fragmentation problem of traditional cascade architectures, yet existing approaches still suffer from flat semantic representations lacking hierarchical structure for multi-step reasoning and an externally constructed chain-of-thought (CoT) that requires expensive annotations and remains disconnected from the generation objective. We propose HoloRec, an endogenous chain-of-thought recommendation mechanism that unifies representation, reasoning, and generation by constructing a hierarchical semantic encoding matrix via multi-granularity nested residual quantization optimized by a holistic reconstruction loss. HoloRec supports two inference modes: a non-thinking mode that uses lightweight multi-granularity supervised alignment for fast prediction, and a thinking mode that employs an interleaved reasoning scheme to generate CoT steps on the fly, directly embedding reasoning into the generation process without external data. Experiments on multiple public recommendation datasets demonstrate that HoloRec consistently outperforms   baselines, with especially significant gains in sparse scenarios, and the thinking mode achieves better accuracy than the non-thinking mode with only modest inference overhead.

\end{abstract}

\ccsdesc[500]{Do Not Use This Code~Generate the Correct Terms for Your Paper}
\ccsdesc[300]{Do Not Use This Code~Generate the Correct Terms for Your Paper}
\ccsdesc{Do Not Use This Code~Generate the Correct Terms for Your Paper}
\ccsdesc[100]{Do Not Use This Code~Generate the Correct Terms for Your Paper}


\received{20 February 2007}
\received[revised]{12 March 2009}
\received[accepted]{5 June 2009}

\maketitle

\section{Introduction}
Generative recommendation systems reformulate the recommendation task as sequence generation. They leverage the semantic understanding and autoregressive capabilities of large language models to produce recommendations in an end-to-end manner \cite{li2024large,deng2025onerec,zhou2026openonerectechnicalreport}. This paradigm overcomes the objective fragmentation and error accumulation problems inherent in traditional cascade architectures that consist of recall, coarse ranking, and fine ranking stages \cite{covington2016deep,zhai2024actions}. It has become a promising research direction because it unifies diverse recommendation signals and naturally enables the incorporation of world knowledge and complex reasoning\cite{yi2025recgpt}.

Despite these advances, existing generative recommendation methods still face three fundamental challenges. Methods such as TIGER \cite{rajput2023recommender} quantize item identifiers into semantic tokens, while P5 \cite{geng2022recommendation} and InstructRec \cite{zhang2026recommendation} transform various recommendation tasks into text generation. However, these approaches suffer from the following limitations.

First, item semantic representations are typically flat vectors or flat token sequences\cite{cheng2026capsidsoftroutedvariablelengthsemantic,khrylchenko2026variablelengthsemanticidsrecommender}. They lack a hierarchical structure that can provide coarse-to-fine granularity information explicitly for multi-step reasoning\cite{wei2025cofireccoarsetofinetokenizationgenerative}. Second, constructing high quality chain-of-thought data for recommendation relies heavily on manual annotation\cite{jiang2026scoterstructuredchainofthoughttransfer}. This leads to high acquisition costs and difficulty in scaling, resulting in extreme scarcity of such data \cite{pang2025reasoning}. Third, even when the chain-of-thought data is available, existing approaches often separate the generation of reasoning steps from the final item prediction into disjoint stages \cite{wang2024interpret,jiang2026endtoendsemanticidgeneration,su2026quantizedinferenceonerecv2}. This separation prevents the reasoning process from deeply guiding the generation process.

To address these challenges, we propose HoloRec, an endogenous chain-of-thought recommendation mechanism that unifies representation, reasoning, and generation. The main contributions of this paper are summarized as follows.

\begin{enumerate}
    \item \textbf{Holistic semantic encoding with hierarchical structure.} We introduce multi-granularity nested residual quantization combined with a holistic reconstruction loss to construct a hierarchical semantic encoding matrix. This matrix provides coarse-to-fine representations that are inherently suitable for reasoning. Moreover, the reconstruction objective generates self-supervised signals that mitigate the scarcity of chain-of-thought data.
    
    \item \textbf{Dual inference modes for endogenous reasoning.} We design two complementary modes within a single model. The non-thinking mode employs lightweight multi-granularity supervised alignment to improve recommendation accuracy without additional inference overhead. The thinking mode uses a coarse-to-fine interleaved reasoning mechanism that injects coarse semantic embeddings into fine-grained decoding via a gating mechanism. This embeds the chain of thought directly into autoregressive generation, thereby bridging the gap between reasoning and recommendation generation.
    
    \item  Experiments on multiple public recommendation datasets demonstrate that HoloRec consistently outperforms strong baselines. The gains are particularly significant on sparse datasets. The thinking mode achieves better accuracy than the non-thinking mode with only modest inference overhead.  
\end{enumerate}

\section{Related Work}
\label{sec:related work}
\subsection{Item Tokenization Representations}
The first step in generative recommendation is to convert items into discrete token sequences. Existing methods \cite{var-len-sid,deng2025onerec,rafailov2023direct,wang2025learnableitemtokenizationgenerative} can be broadly divided into two categories. The first category builds on language models and textual descriptions, directly using item metadata such as titles and descriptions and leveraging the semantic understanding capabilities of pretrained language models to handle recommendation tasks\cite{tan2024idgenrecllmrecsysalignmenttextual}. These methods do not require vocabulary expansion, but the text lengths are inconsistent, efficiency is relatively low, and long texts tend to introduce redundancy. The second category relies on semantic IDs, obtaining compact discrete codes by quantizing continuous semantic embeddings\cite{ju2025generative,cheng2026capsidsoftroutedvariablelengthsemantic,ju2025generativerecommendationsemanticids}. A representative work, Residual Quantized Variational AutoEncoder (RQ-VAE)\cite{rajput2023recommender}, applies residual quantization to item embeddings layer by layer and concatenates the codebook indices from each layer into a semantic ID. Subsequently, RQ-KMeans\cite{deng2025onerec} replaces end-to-end training with discrete clustering, improving efficiency and stability. The above methods all target the reconstruction of original semantics, resulting in flat encoding structures that lack the hierarchical information needed to support coarse-to-fine reasoning\cite{wei2025cofireccoarsetofinetokenizationgenerative,jiang2026endtoendsemanticidgeneration}.

\subsection{Generative Recommendation}   
After obtaining tokenized representations of items, the model needs to predict the next item from the user's historical sequence. Existing modeling approaches fall into autoregressive generation and parallel generation. Autoregressive generation treats recommendation as a standard sequence generation task and predicts token by token based on encoder-decoder or decoder-only architectures. TIGER\cite{rajput2023recommender} first represents items as semantic IDs, converts interaction sequences into semantic ID sequences, and performs sequence-to-sequence generation via a Transformer. OneRec\cite{deng2025onerec} further introduces collaborative similarity alignment to enhance generation accuracy. Parallel generation aims to improve decoding efficiency. NAR4Rec\cite{ren2024non} adopts a non-autoregressive generator to output recommendation lists in parallel, and LLaDA-Rec\cite{shi2025llada} replaces the autoregressive model with a discrete diffusion model, mitigating error accumulation through bidirectional generation. Parallel methods offer clear advantages in inference latency, yet they struggle to adequately model semantic dependencies among tokens, leaving a trade-off between accuracy and efficiency.

\begin{figure*}[tp]
    \centering  
    \includegraphics[width=0.75\textwidth]{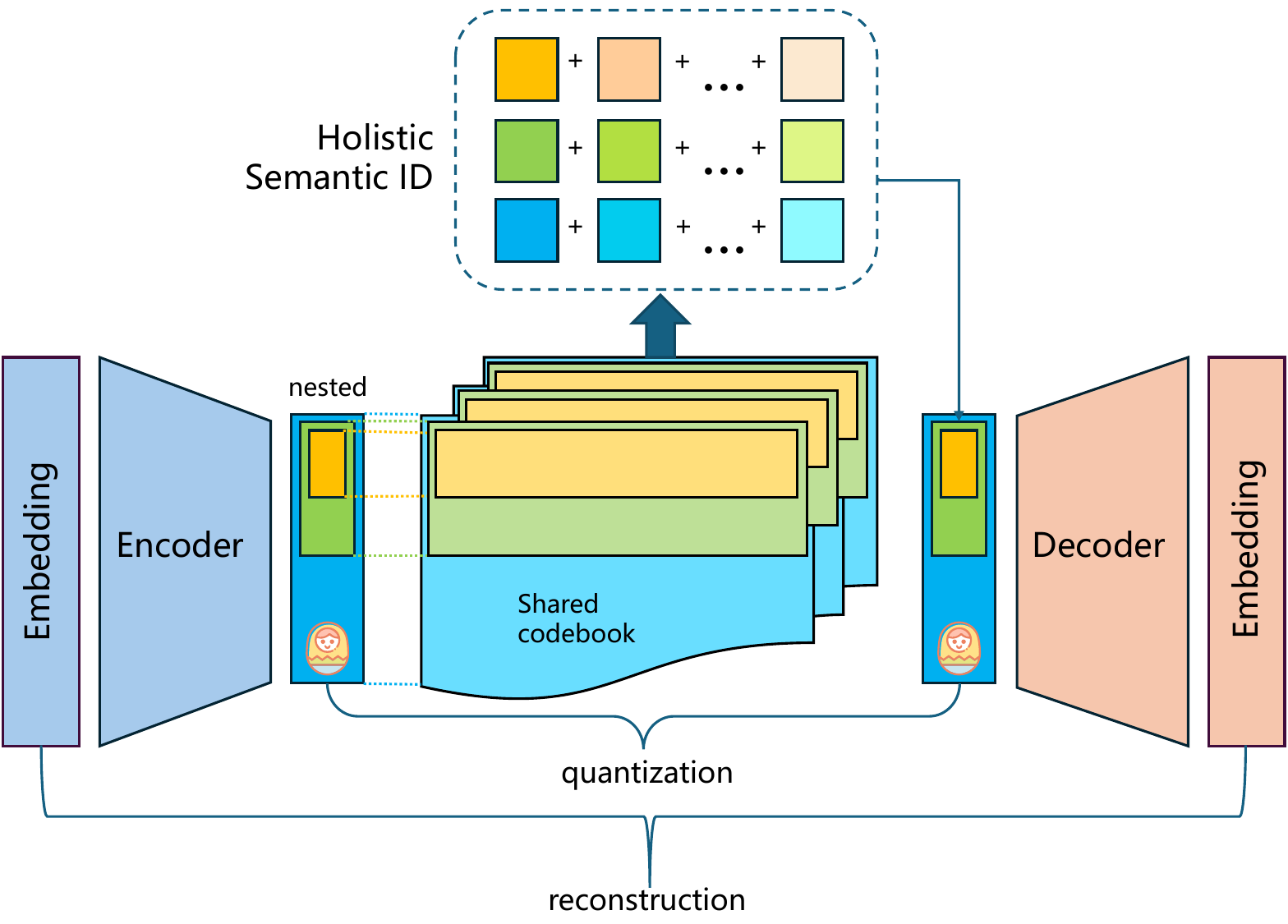}
    \caption{\textbf{Structure of Holistic Semantic Encoding}}
    \label{fig:holistic_encoding}
\end{figure*}

\subsection{Reasoning-Enhanced Recommendation }   
Inspired by chain-of-thought techniques, recommendation systems have introduced reasoning to better capture user intent\cite{yao2023tree,lightman2023let,lin2026verifiablereasoningllmbasedgenerative,you2025r2eclargerecommendermodels}. Existing studies can be categorized into explicit reasoning and implicit reasoning. Explicit reasoning generates readable natural language as intermediate reasoning steps\cite{Lei_2024,yang2025reccot,jiang2026scoterstructuredchainofthoughttransfer}. CoT4Rec\cite{yue2025cot4rec} constructs zero-shot prompts by clustering sequences and selecting exemplars, using chain-of-thought to guide generation. OneRec-Think\cite{liu2025onerec} employs natural language to reason about user preferences and item relationships before generation. R2Rec\cite{zhao2025reason} samples chain structures from interaction graphs and transforms them into interactive reasoning. Explicit reasoning offers strong interpretability but suffers from the scarcity of high-quality chain-of-thought data and high inference latency\cite{meng2025keragrknowledgeenhancedretrievalaugmentedgeneration}. Implicit reasoning discards explicit natural language and internalizes the reasoning process in the hidden representation space. ReaRec\cite{tang2025think} introduces a test-time computation framework, decoupling the encoding and reasoning spaces through autoregressive hidden states and special position embeddings. REG4Rec\cite{xing2025reg4rec} employs a mixture of experts and parallel quantization to generate diverse semantic representations, enhanced by reinforcement learning for reliability. OnePiece\cite{dai2025onepiece} integrates context engineering with block-wise implicit reasoning and uses progressive multi-objective training to strengthen supervision. Implicit reasoning is more efficient, but the reasoning process typically lacks explicit interpretability\cite{li2025implicit}, and the reasoning and recommendation generation modules often remain mutually independent, failing to achieve deep integration.

\section{Approach}
\label{sec:approach}
The proposed endogenous chain-of-thought recommendation mechanism, named HoloRec, comprises three progressively integrated modules. First, a holistic semantic encoding module constructs hierarchical item representations. Second, a non-thinking mode with multi-granularity supervised alignment training introduces lightweight hierarchical supervision. Third, an endogenous chain-of-thought model based on interleaved reasoning integrates coarse-to-fine reasoning into the sequence generation process. Figure \ref{fig:holistic_encoding} illustrates the general framework.

\subsection{Holistic Semantic Encoding}
Holistic semantic encoding generates an \(M \times L\) discrete encoding matrix for each item, where \(M\) denotes the number of semantic granularity levels and \(L\) denotes the number of residual quantization layers. The rows of the matrix from top to bottom correspond to coarse to fine semantic levels, for example 8, 16, and 32 dimensions, while the columns from left to right correspond to progressively refined residual quantization layers.

\subsubsection{Constructing the hierarchical matrix}

To construct this matrix, we first obtain a latent representation of the item. Given the original embedding \(\mathbf{e}\) of an item, the encoder outputs a latent vector
\(\mathbf{z} \in \mathbb{R}^D.\)  We then define a dimension sequence 
\(d_1 < d_2 < \dots < d_M = D\) 
extract the first dimensions \(d_m\) of \(\mathbf{z}\) as the initial representation at the \(m\)th granularity, denoted \(\mathbf{z}^{(0)}_m\). 
Subsequently, for each granularity, we independently perform \(L\) layers of residual quantization. All granularities share the codebook at each layer \(l\):
\(\mathbf{C}_l = \{\mathbf{c}_{l,1}, \dots, \mathbf{c}_{l,K}\}.
\)

\subsubsection{Training Process}

Given the initial residual representation \(\mathbf{r}^{(0)}_m = \mathbf{z}^{(0)}_m\), the quantization process at granularity \(m\) and layer \(l\) proceeds in three steps: selecting the nearest codebook vector, recording the quantized vector, and updating the residual for the next layer. Formally,

\begin{equation}
    \begin{aligned}
        q_{m,l} &= \arg\min_{k} \|\mathbf{r}^{(l-1)}_m - \mathbf{c}_{l,k[:d_m]}\|_2^2,\\
        \hat{\mathbf{r}}^{(l)}_m &= \mathbf{c}_{l,q_{m,l}},\\
        \mathbf{r}^{(l)}_m &= \mathbf{r}^{(l-1)}_m - \hat{\mathbf{r}}^{(l)}_m[:d_m].
    \end{aligned}
\end{equation}
Here, \(\mathbf{r}^{(l-1)}_m\) is the residual before layer \(l\), and \(\mathbf{c}_{l,k[:d_m]}\) denotes the first \(d_m\) dimensions of the \(k\) th code vector in codebook \(\mathbf{C}_l\). After processing all \(L\) layers, the cumulative quantized vector for granularity \(m\) is \(\hat{\mathbf{z}}_m = \sum_{l=1}^L \hat{\mathbf{r}}^{(l)}_m\). The indices \(q_{m,l}\) across all \(m\) and \(l\) are then stacked into the holistic semantic matrix.

The training objective consists of three complementary loss functions.

\textbf{Reconstruction loss.}
This loss ensures that the finest granularity encoding can recover the original item embedding:
\begin{equation}
    \mathcal{L}_{\text{recon}} = \|\mathrm{Dec}(\hat{\mathbf{z}}_M) - \mathbf{e}\|_2^2,
\end{equation}
where \(\mathrm{Dec}\) is the decoder and \(\mathbf{e}\) is the original embedding.

\textbf{Quantization loss.}
This loss supervises the learning of all codebooks:
\begin{equation}
\mathcal{L}_{\text{quant}} = \frac{1}{LM}\sum_{l=1}^{L}\sum_{m=1}^{M} \|\mathbf{r}^{(l-1)}_m - \hat{\mathbf{r}}^{(l)}_m[:d_m]\|_2^2.
\end{equation}

\textbf{Holistic reconstruction loss.}
This loss enforces that the cumulative quantized vector at each granularity can independently reconstruct its corresponding initial slice without relying on finer details:
\begin{equation}
\mathcal{L}_{\text{holo}} = \frac{1}{M}\sum_{m=1}^{M} \|\hat{\mathbf{z}}_m - \mathbf{z}^{(0)}_m\|_2^2.
\end{equation}

The total loss is a weighted combination:
\begin{equation}
\mathcal{L} = \mathcal{L}_{\text{recon}} + \lambda_1 \mathcal{L}_{\text{quant}} + \lambda_2 \mathcal{L}_{\text{holo}},
\end{equation}
where \(\lambda_1\) and \(\lambda_2\) are hyperparameters that balance the contribution of the quantization loss and the holistic reconstruction loss, respectively. 

This design produces a hierarchical encoding matrix where entries become progressively finer from the upper left to the lower right. Moreover, the holistic reconstruction loss provides multi-granularity supervision signals that benefit subsequent reasoning tasks. More details are shown in the algorithm \ref{alg:holographic_coding}.

\subsubsection{Comparison with conventional residual quantization.}
This design differs from standard residual quantization methods such as RQ-VAE, which applies a single granularity of representation and performs residual quantization only on the full embedding. In contrast, our holistic encoding produces a multi-granularity matrix that captures semantic information from coarse to fine levels. This hierarchical structure explicitly supports downstream reasoning tasks by providing intermediate semantic abstractions.

Moreover, sharing codebooks across granularities reduces parameter overhead while preserving the distinctness of each quantization layer. The holistic reconstruction loss introduced later further enforces that each granularity independently reconstructs the original semantics, a property absent in traditional RQ-VAE.

\begin{algorithm}[tp]
\caption{Holographic Semantic Encoding Training}
\label{alg:holographic_coding}
\SetAlgoLined
\KwIn{Item embedding set $\{\mathbf{e}_i\}_{i=1}^N$, number of granularities $M$, number of residual quantization layers $L$, shared codebooks $\mathcal{C} = \{\mathbf{c}_1, \ldots, \mathbf{c}_K\}$, hyperparameters $\lambda_1, \lambda_2$, batch size $B$, learning rate $\eta$.}
\KwOut{Trained encoder Enc, decoder Dec, shared codebooks $\mathcal{C}$.}

Initialize $\mathcal{C}$ via K-Means clustering\;

\For{$epoch = 1$ to $E$}{
    \For{each batch $\{\mathbf{e}_i\}_{i=1}^B$}{
        $\mathbf{z}_i = \text{Enc}(\mathbf{e}_i)$\;
        Extract $M$ nested sub-vectors $\mathbf{z}_{i,m}^{(0)} = \mathbf{z}_i[:d_m]$ for $m=1,\dots,M$, where $d_1 < \dots < d_M = D$\;
        
        \For{$m = 1$ to $M$}{
            $\mathbf{r}_{i,m}^{(0)} \gets \mathbf{z}_{i,m}^{(0)}$, $\hat{\mathbf{z}}_{i,m} \gets \mathbf{0}$\;
            \For{$l = 1$ to $L$}{
                $q_{i,m,l} \gets \arg\min_{k} \|\mathbf{r}_{i,m}^{(l-1)} - \mathbf{c}_k[:d_m]\|_2^2$\;
                $\hat{\mathbf{r}}_{i,m}^{(l)} \gets \mathbf{c}_{q_{i,m,l}}$\;
                $\hat{\mathbf{z}}_{i,m} \gets \hat{\mathbf{z}}_{i,m} + \hat{\mathbf{r}}_{i,m}^{(l)}$\;
                $\mathbf{r}_{i,m}^{(l)} \gets \mathbf{r}_{i,m}^{(l-1)} - \hat{\mathbf{r}}_{i,m}^{(l)}[:d_m]$\;
            }
            Obtain code index sequence $[q_{i,m,1}, \dots, q_{i,m,L}]$\;
        }
        
        Assemble holographic semantic matrix $\mathbf{Q}_i \in \mathbb{N}^{M \times L}$\;
        $\hat{\mathbf{e}}_i = \text{Dec}(\hat{\mathbf{z}}_{i,M})$\;
        
        Compute reconstruction loss:
        $\mathcal{L}_{\text{recon}} = \frac{1}{B} \sum_i \|\hat{\mathbf{e}}_i - \mathbf{e}_i\|_2^2$\;
        
        Compute quantization loss:
        $\mathcal{L}_{\text{quant}} = \frac{1}{B L M} \sum_{i,m,l} \|\mathbf{r}_{i,m}^{(l-1)} - \hat{\mathbf{r}}_{i,m}^{(l)}[:d_m]\|_2^2$\;
        
        Compute holistic reconstruction loss:
        $\mathcal{L}_{\text{holo}} = \frac{1}{B M} \sum_{i,m} \|\hat{\mathbf{z}}_{i,m} - \mathbf{z}_{i,m}^{(0)}\|_2^2$\;
        
        $\mathcal{L} = \mathcal{L}_{\text{recon}} + \lambda_1 \mathcal{L}_{\text{quant}} + \lambda_2 \mathcal{L}_{\text{holo}}$\;
        Perform backpropagation and update Enc, Dec, $\mathcal{C}$ with learning rate $\eta$\;
    }
}
\Return Enc, Dec, $\mathcal{C}$\;
\end{algorithm}

\subsubsection{Complexity analysis}
The offline training complexity of holistic semantic encoding is \(O(M L K D)\), which is \(M\) times that of standard RQVAE with \(O(L K D)\). The parameter overhead scales as \(O(M D^2)\) compared to the base encoder decoder. Both increases are incurred only once during offline preprocessing and do not affect online inference complexity.

\begin{figure*}[tp]
    \centering
    \includegraphics[width=0.75\textwidth]{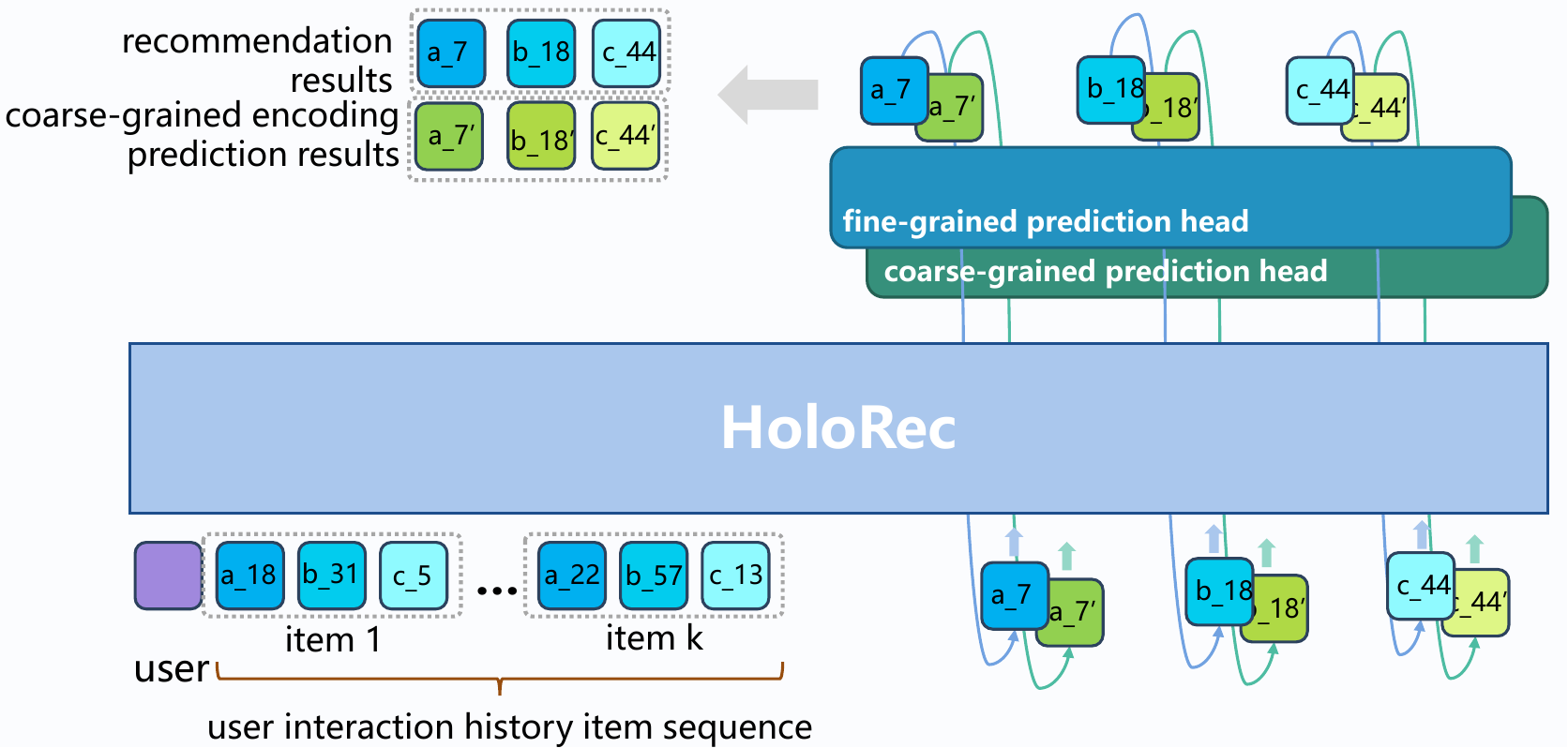}
    \Description{Model structure with fine and coarse prediction heads.}
    \caption{Model structure for multi-granularity supervised alignment training. The fine-grained prediction head outputs the final recommendation, while the coarse-grained head provides hierarchical supervision only during training.}
    \label{fig:non-thinking}
\end{figure*}

\subsection{Non-thinking Mode with Multi-Granularity Supervised Alignment Training}

The non-thinking mode introduces hierarchical supervision into the autoregressive generation process without altering the inference pipeline or adding runtime overhead. Each item in a user history sequence is represented by its finest granularity encoding from the holistic semantic matrix. In addition to the original fine-grained prediction head, we add one or more coarse-grained prediction heads with identical structure but output dimensions matching the corresponding coarse-grained vocabularies. The coarse-grained heads are used only during training and are removed at inference time, preserving efficiency.

\subsubsection{Training Process}

Let \(h_1, \ldots, h_T\) denote the decoder hidden states for a sequence of length \(T\). For each time step \(t\), let \(y_t^{\text{fine}}\) and \(y_t^{\text{coarse}}\) denote the ground truth token indices at the fine and coarse granularities, respectively, obtained from the holistic semantic matrix. Let \(\hat{y}_t^{\text{fine}}(k)\) and \(\hat{y}_t^{\text{coarse}}(k)\) be the predicted probabilities for token \(k\) from the corresponding softmax output.

The training objective consists of two loss functions.

\textbf{Fine-grained cross-entropy loss.} This loss supervises the final recommendation:
\[
\mathcal{L}_{\text{fine}}^{\text{seq}} = -\frac{1}{T} \sum_{t=1}^{T} \log \hat{y}_t^{\text{fine}}(y_t^{\text{fine}}).
\]

\textbf{Coarse-grained cross-entropy loss.} This loss provides hierarchical supervision:
\[
\mathcal{L}_{\text{coarse}}^{\text{seq}} = -\frac{1}{T} \sum_{t=1}^{T} \log \hat{y}_t^{\text{coarse}}(y_t^{\text{coarse}}).
\]

For a mini-batch of \(B\) sequences, the overall losses are averaged over the batch:
\[
\mathcal{L}_{\text{fine}} = \frac{1}{B} \sum_{i=1}^{B} \mathcal{L}_{\text{fine}}^{\text{seq}(i)}, \qquad
\mathcal{L}_{\text{coarse}} = \frac{1}{B} \sum_{i=1}^{B} \mathcal{L}_{\text{coarse}}^{\text{seq}(i)}.
\]

The total loss is a weighted combination:
\[
\mathcal{L}_{\text{total}} = \alpha \mathcal{L}_{\text{fine}} + (1 - \alpha) \mathcal{L}_{\text{coarse}},
\]
where \(\alpha \in (0,1)\) balances the two supervision signals. The coarse-grained ground truth labels are obtained by taking the corresponding row index from the holistic semantic matrix of each item. For instance, if \(M=3\) and the granularity dimensions are \(\{8,16,32\}\), the label for the coarsest granularity (8‑dim) is the first row of the matrix.

All coarse-grained prediction heads are removed at inference time, so the model performs standard autoregressive generation using only the fine-grained head. No extra computation is introduced.
Algorithm~\ref{alg:supervised_alignment} summarizes the training procedure. Figure~\ref{fig:non-thinking} illustrates the model structure with two prediction heads.

\begin{algorithm}[tp]
\caption{Multi-Granularity Supervised Alignment Training}
\label{alg:supervised_alignment}
\SetAlgoLined
\KwIn{User history sequences $\mathcal{X} = \{x_1, \ldots, x_T\}$, fine-grained labels $y^{\text{fine}}$, coarse-grained labels $y^{\text{coarse}}$, encoder Enc, decoder Dec, fine-grained head $P_{\text{fine}}$, coarse-grained head $P_{\text{coarse}}$, hyperparameter $\alpha$, batch size $B$, learning rate $\eta$.}
\KwOut{Trained model parameters $\theta = \{\text{Enc}, \text{Dec}, P_{\text{fine}}, P_{\text{coarse}}\}$.}
Initialize $\theta$\;
\For{each epoch}{
    \For{each batch of $B$ sequences}{
        Encode each sequence $\mathcal{X}_i$ into hidden states $\mathbf{H}_i^{\text{enc}}$ via Enc\;
        Decode autoregressively to obtain hidden states $\mathbf{h}_{i,1}^{\text{dec}}, \ldots, \mathbf{h}_{i,T}^{\text{dec}}$\;
        \For{each position $t$}{
            Compute $\hat{y}_{i,t}^{\text{fine}} = P_{\text{fine}}(\mathbf{h}_{i,t}^{\text{dec}})$, $\hat{y}_{i,t}^{\text{coarse}} = P_{\text{coarse}}(\mathbf{h}_{i,t}^{\text{dec}})$\;
        }
        Compute $\mathcal{L}_{\text{fine}} = -\frac{1}{B T} \sum_{i=1}^B \sum_{t=1}^T \log \hat{y}_{i,t}^{\text{fine}}(y_{i,t}^{\text{fine}})$\;
        Compute $\mathcal{L}_{\text{coarse}} = -\frac{1}{B T} \sum_{i=1}^B \sum_{t=1}^T \log \hat{y}_{i,t}^{\text{coarse}}(y_{i,t}^{\text{coarse}})$\;
        $\mathcal{L}_{\text{total}} = \alpha \mathcal{L}_{\text{fine}} + (1-\alpha) \mathcal{L}_{\text{coarse}}$\;
        Update $\theta$ using gradient descent\;
    }
}
\Return $\theta$\;
\end{algorithm}

\begin{figure*}[t]
    \centering
    \includegraphics[width=0.8\textwidth]{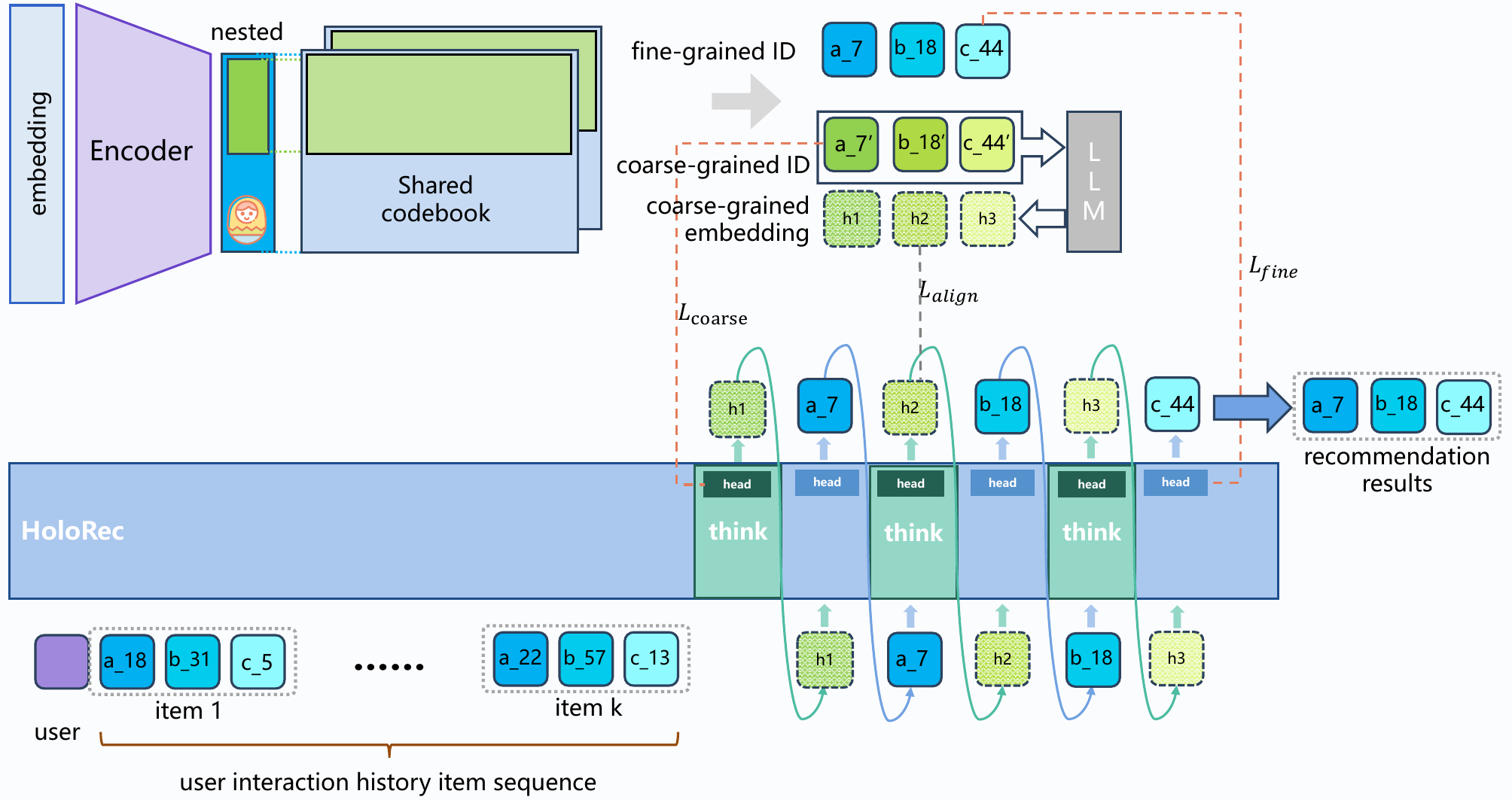}
    \caption{Structure of the interleaved reasoning endogenous chain-of-thought model. At each decoding step, the fine-grained hidden state is first used to predict a coarse-grained semantic embedding, which is then fused with the decoder contextual state via a gating injection module. The resulting enhanced fine-grained representation is passed to the fine-grained prediction head to output the final recommendation.}
    \label{fig:holorec}
\end{figure*}

\subsection{Interleaved Reasoning Endogenous Chain of Thought Model}

The thinking mode integrates coarse-to-fine reasoning directly into the autoregressive generation process. At each decoding step, the model first predicts a coarse-grained semantic distribution based on the previous fine-grained hidden state. This coarse information is then transformed into a soft embedding and fused with the decoder contextual state via a gating mechanism. The enhanced fine-grained hidden state is finally used for fine-grained token prediction.

\subsubsection{Training Process}

Let \(\mathbf{h}_{t-1}^{\text{fine}}\) denote the fine-grained hidden state from the previous step. The coarse-grained distribution at step \(t\) is computed as
\begin{equation}
    \mathbf{p}_t^{\text{coarse}} = \mathrm{Softmax}\bigl( \mathbf{W}_c \mathbf{h}_{t-1}^{\text{fine}} \bigr),
\end{equation}
where \(\mathbf{W}_c\) is a learnable projection matrix. This distribution is used to obtain a soft coarse embedding via a weighted sum over the coarse embedding matrix \(\mathbf{E}\):
\begin{equation}
    \tilde{\mathbf{e}}_t = \sum_{i} p_{t,i}^{\text{coarse}} \cdot \mathbf{E}_i.
\end{equation}

The soft embedding \(\tilde{\mathbf{e}}_t\) serves as an additional input to the decoder. A forward pass produces the contextual hidden state \(\mathbf{h}_t^{\text{ctx}}\):
\begin{equation}
    \mathbf{h}_t^{\text{ctx}} = \mathrm{Dec}\bigl( \tilde{\mathbf{e}}_t,\ \mathbf{H}^{\text{enc}} \bigr),
\end{equation}
where \(\mathbf{H}^{\text{enc}}\) is the encoded user history sequence.

A gating mechanism fuses the coarse information into the fine grained representation. The gate coefficients and a candidate correction are computed as
\begin{equation}
    \mathbf{g}_t = \sigma\bigl( \mathbf{W}_g [\mathbf{h}_t^{\text{ctx}}; \tilde{\mathbf{e}}_t] + \mathbf{b}_g \bigr), \qquad
    \Delta \mathbf{h}_t = \mathbf{W}_p \tilde{\mathbf{e}}_t,
\end{equation}
where \(\sigma\) is the sigmoid function, \(\mathbf{W}_g\) and \(\mathbf{W}_p\) are weight matrices, and \(\mathbf{b}_g\) is a bias. The enhanced fine-grained hidden state is obtained via a residual connection followed by layer normalization:
\begin{equation}
    \mathbf{h}_t^{\text{fine}} = \mathrm{LayerNorm}\bigl( \mathbf{h}_t^{\text{ctx}} + \mathbf{g}_t \odot \Delta \mathbf{h}_t \bigr).
\end{equation}

Finally, the fine-grained prediction head produces the recommendation token based on \(\mathbf{h}_t^{\text{fine}}\). 

\textbf{Coarse-grained cross-entropy loss.} This loss supervises the coarse-grained predictions:
\begin{equation}
    \mathcal{L}_{\text{coarse}}=-\frac{1}{T}\sum_{i=1}^T\log \mathbf{p}_t^{coarse}(y_t^{coarse}),
\end{equation}
where \(y_t^{coarse}\) is the ground truth coarse token at step \(t\).
\textbf{Fine-grained cross-entropy loss.} This loss supervises the final recommendation:
\begin{equation}
    \mathcal{L}_{\text{fine}}=-\frac{1}{T}\sum_{i=1}^T\log \mathbf{p}_t^{fine}(y_t^{fine}),
\end{equation}
where \(y_t^{fine}\) is the ground truth fine token at step \(t\) and \(\mathbf{p}_t^{fine}\) is the output distribution from the fine-grained head.
\textbf{Alignment loss.} This loss encourages the soft embedding to stay close to the ground truth coarse embedding:
\begin{equation}
    \mathcal{L}_{align}=\frac{1}{T}\sum_{i=1}^T\bigl\| \tilde{\mathbf{e}}_t - \mathbf{E}_{y_t^{\text{coarse}}} \bigr\|_2^2,
\end{equation}
where \(\mathbf{E}_{y_t^{\text{coarse}}}\) is the embedding of the ground truth coarse token.
The total loss is a weighted combination:
\begin{equation}
    \mathcal{L}_{\text{total}} = \lambda_1 \mathcal{L}_{\text{coarse}} + \lambda_2 \mathcal{L}_{\text{fine}} + \lambda_3 \mathcal{L}_{\text{align}},
\end{equation}
where \(\lambda_1,\lambda_2, \lambda_3\) are hyperparameters that balance the contributions of the three losses.

Teacher forcing is used during training, where the decoder state is updated with the true fine-grained embedding. At inference time, the same interleaved steps are executed autoregressively without the alignment loss. Algorithm~\ref{alg:interleaved_reasoning} summarizes the training procedure, and Figure~\ref{fig:holorec} illustrates the overall model architecture.

\begin{algorithm}[tp]
\caption{Interleaved Reasoning with Endogenous Chain of Thought Training}
\label{alg:interleaved_reasoning}
\SetAlgoLined
\KwIn{User history sequence $\mathcal{X}$, coarse labels $y^{\text{coarse}}$, fine labels $y^{\text{fine}}$; encoder Enc, decoder Dec; coarse head $P_{\text{coarse}}$, fine head $P_{\text{fine}}$; coarse embedding matrix $\mathbf{E}$; gate parameters $\mathbf{W}_g, \mathbf{b}_g, \mathbf{W}_p$; hyperparameters $\lambda_1, \lambda_2, \lambda_3$, learning rate $\eta$.}
\KwOut{Trained model parameters $\theta$.}
Initialize $\theta$\;
\For{each epoch}{
    \For{each batch $(\mathcal{X}, y^{\text{coarse}}, y^{\text{fine}})$}{
        $\mathbf{H}^{\text{enc}} = \text{Enc}(\mathcal{X})$\;
        $\mathbf{h}_0^{\text{fine}} \gets \mathbf{0}$\;
        $\mathcal{L}_{\text{coarse}} \gets 0$, $\mathcal{L}_{\text{fine}} \gets 0$, $\mathcal{L}_{\text{align}} \gets 0$\;
        \For{$t = 1$ to $T$}{
            $\mathbf{p}_t^{\text{coarse}} = \mathrm{Softmax}(P_{\text{coarse}}(\mathbf{h}_{t-1}^{\text{fine}}))$\;
            $\tilde{\mathbf{e}}_t = \sum_i p_{t,i}^{\text{coarse}} \cdot \mathbf{E}_i$\;
            $\mathbf{h}_t^{\text{ctx}} = \text{Dec}(\tilde{\mathbf{e}}_t, \mathbf{H}^{\text{enc}})$\;
            $\mathbf{g}_t = \sigma(\mathbf{W}_g[\mathbf{h}_t^{\text{ctx}}; \tilde{\mathbf{e}}_t] + \mathbf{b}_g)$\;
            $\Delta \mathbf{h}_t = \mathbf{W}_p \tilde{\mathbf{e}}_t$\;
            $\mathbf{h}_t^{\text{fine}} = \mathrm{LayerNorm}(\mathbf{h}_t^{\text{ctx}} + \mathbf{g}_t \odot \Delta \mathbf{h}_t)$\;
            $\mathbf{p}_t^{\text{fine}} = \mathrm{Softmax}(P_{\text{fine}}(\mathbf{h}_t^{\text{fine}}))$\;
            $\mathcal{L}_{\text{fine}} \leftarrow \mathcal{L}_{\text{fine}} - \log \mathbf{p}_t^{\text{fine}}(y_t^{\text{fine}})$\;
            $\mathcal{L}_{\text{coarse}} \leftarrow \mathcal{L}_{\text{coarse}} - \log \mathbf{p}_t^{\text{coarse}}(y_t^{\text{coarse}})$\;
            $\mathcal{L}_{\text{align}} \leftarrow \mathcal{L}_{\text{align}} + \|\tilde{\mathbf{e}}_t - \mathbf{E}_{y_t^{\text{coarse}}}\|_2^2$\;
        }
        $\mathcal{L}_{\text{total}} = \lambda_1 \mathcal{L}_{\text{coarse}} + \lambda_2 \mathcal{L}_{\text{fine}} + \lambda_3 \mathcal{L}_{\text{align}}$\;
        Update $\theta$ using gradient descent\;
    }
}
\Return $\theta$\;
\end{algorithm}

\subsubsection{Complexity Analysis}
Compared with the non thinking mode, the thinking mode adds one extra decoder forward pass per step for each auxiliary granularity to obtain \(\mathbf{h}_t^{\text{ctx}}\). When employing all \(M-1\) auxiliary granularities, the constant factor increases roughly \(M\) times. The asymptotic time complexity becomes \(O(MT^2 d_{\text{model}})\), where \(T\) is the sequence length and \(d_{\text{model}}\) is the hidden dimension. The gating and injection modules introduce two linear layers, adding \(O(d_{\text{model}}^2)\) parameters, which is negligible relative to the backbone. Memory footprint is nearly identical to the baseline.

\section{Experiments}
\label{sec:exp}
In this section, we conduct comprehensive experiments to evaluate the proposed HoloRec framework. We aim to answer the following research questions:
\begin{itemize}
    \item \textbf{RQ1}: How does HoloRec compare with state-of-the-art baselines on public recommendation datasets?
    \item \textbf{RQ2}: How do the holistic semantic encoding, multi-granularity supervised alignment, and interleaved reasoning mechanism contribute to the overall performance?
    \item \textbf{RQ3}: How does the proposed method perform under different sequence lengths?
    \item \textbf{RQ4}: How do different granularity encodings affect recommendation performance?
    \item \textbf{RQ5}: How does the supervision task weight hyperparameter influence the multi-granularity alignment?
\end{itemize}
 
\subsection{Experimental Setup}

\subsubsection{Datasets}
Following previous studies \cite{rajput2023recommender, kang2018self, sun2019bert4rec, deng2025onerec}, we evaluate our approach on two public datasets of different scales: Beauty and Instruments\cite{ni2019justifying}. We apply five-core filtering to discard users and items with fewer than five interactions, and adopt the leave-one-out strategy for data splitting. The statistics of the datasets after preprocessing are summarized in Table~\ref{tab:dataset_stats}.

\begin{table}[tp]
    \centering
    \caption{Dataset Statistics.}
    \label{tab:dataset_stats}
    \footnotesize
    \setlength{\tabcolsep}{6pt}
    \renewcommand{\arraystretch}{1.2}
    \begin{tabular}{lcccc}
        \toprule
        \textbf{Dataset} & \textbf{\#Users} & \textbf{\#Items} & \textbf{\#Interactions} & \textbf{Density} \\
        \midrule
        Beauty      & 22,363 & 12,100 & 198,502 & 0.073\% \\
        Instruments & 24,772 &  9,921 & 206,153 & 0.084\% \\
        \bottomrule
    \end{tabular}
\end{table}


\subsubsection{Metrics}
For evaluation, we employ Hit Rate (Hit@N) and Normalized Discounted Cumulative Gain (NDCG@N) with \(N \in \{5, 10\}\). Hit@N measures whether the ground truth item appears in the top-N recommendation list, while NDCG@N further accounts for the ranking positions of the correctly recommended items.

\begin{table*}[tp]
    \caption{Overall Performance Comparison}
    \label{tab:main}
    \centering
    \renewcommand{\arraystretch}{1.2}
    \begin{tabular}{lcccccccc}
        \toprule
        \multirow{2}{*}{\textbf{Model}} & \multicolumn{4}{c}{\textbf{Beauty}} & \multicolumn{4}{c}{\textbf{Instruments}} \\
        \cmidrule(lr){2-5} \cmidrule(lr){6-9}
        & \textbf{Hit@5} & \textbf{Hit@10} & \textbf{NDCG@5} & \textbf{NDCG@10} & \textbf{Hit@5} & \textbf{Hit@10} & \textbf{NDCG@5} & \textbf{NDCG@10} \\
        \midrule
        MF      & 0.0294 & 0.0474 & 0.0145 & 0.0191 & 0.0479 & 0.0735 & 0.0330 & 0.0412 \\
        HGN     & 0.0325 & 0.0512 & 0.0206 & 0.0266 & 0.0813 & 0.1048 & 0.0668 & 0.0774 \\
        BERT4Rec& 0.0203 & 0.0347 & 0.0124 & 0.0170 & 0.0671 & 0.0822 & 0.0560 & 0.0608 \\
        SASRec  & 0.0380 & 0.0588 & 0.0246 & 0.0313 & 0.0751 & 0.0947 & 0.0627 & 0.0690 \\
        BIGRec  & 0.0243 & 0.0299 & 0.0181 & 0.0198 & 0.0513 & 0.0576 & 0.0470 & 0.0491 \\
        P5-SemID& 0.0393 & 0.0584 & 0.0273 & 0.0335 & 0.0775 & 0.0964 & 0.0669 & 0.0730 \\
        TIGER   & 0.0395 & 0.0610 & 0.0262 & 0.0331 & 0.0870 & 0.1058 & 0.0737 & 0.0797 \\
        \midrule
        \textbf{HoloRec} & \textbf{0.0451} & \textbf{0.0683} & \textbf{0.0307} & \textbf{0.0375} & \textbf{0.0894} & \textbf{0.1105} & \textbf{0.0761} & \textbf{0.0828} \\
        \bottomrule
    \end{tabular}
\end{table*}

\subsubsection{Baselines}
We compare HoloRec against two categories of baselines.

\textbf{Traditional recommendation models.} 
\begin{itemize}
    \item \textbf{MF}: Matrix factorization based on collaborative filtering.
    \item \textbf{HGN}: Hierarchical graph network that captures higher-order user-item interactions.
    \item \textbf{SASRec} \cite{kang2018self}: A unidirectional self-attention based sequential model.
    \item \textbf{BERT4Rec} \cite{sun2019bert4rec}: A bidirectional self-attention model for sequence recommendation.
\end{itemize}

\textbf{Generative recommendation models.}
\begin{itemize}
    \item \textbf{BIGRec} \cite{li2024large}: Uses plain text identifiers for item representation and generates recommendations via language models.
    \item \textbf{P5-SemID} \cite{geng2022recommendation}: Transforms recommendation tasks into text generation with semantic IDs.
    \item \textbf{TIGER} \cite{rajput2023recommender}: Applies residual quantization to generate semantic item identifiers in a sequence-to-sequence manner.
\end{itemize}


\subsubsection{Implementation Details}
To enable a fair comparison with the baseline TIGER, which adopts the T5 architecture, our method also uses T5 as the backbone model. For holistic semantic encoding, we set the number of granularity levels \(M = 3\) with dimensions \(\{8, 16, 32\}\), the number of residual quantization layers \(L = 4\), and codebook size \(K = 256\) for each layer. All codebooks are shared across granularities and initialized via K-Means clustering on the latent representations. The hyperparameters for the total loss are set to \(\lambda_1 = 1.0\) and \(\lambda_2 = 1\). For the non-thinking mode, we set \(\alpha = 0.5\) to balance the fine-grained and coarse-grained cross-entropy losses. For the thinking mode, we set \(\lambda_1 = 5\), \(\lambda_2 = 1\), and \(\lambda_3 = 200\) for the coarse, fine, and alignment losses, respectively. We train all generative models for up to 200 epochs with an early stopping patience of 20. The learning rate is set to \(5 \times 10^{-4}\) with a warmup ratio of 0.1 and cosine decay. The batch size is 1024. All models are implemented in PyTorch and trained on NVIDIA A100 GPUs.

\begin{table*}[tp]
    \caption{Ablation Study}
    \label{tab:method_comparison}
    \centering
    \renewcommand{\arraystretch}{1.2}
    \begin{tabular}{llccccc}
        \toprule
        \textbf{Dataset} & \textbf{Model} & \textbf{Hit@1} & \textbf{Hit@5} & \textbf{Hit@10} & \textbf{NDCG@5} & \textbf{NDCG@10} \\
        \midrule
        \multirow{3}{*}{Beauty} 
        & w/o aux  & 0.0140 & 0.0425 & 0.0661 & 0.0283 & 0.0359 \\
        & w/o Interleaved   & 0.0140 & 0.0431 & 0.0682 & 0.0288 & 0.0369 \\
        & HoloRec & 0.0160 & 0.0451 & 0.0683 & 0.0307 & 0.0375 \\
        \midrule
        \multirow{3}{*}{Instruments} 
        & w/o aux  & 0.0612 & 0.0883 & 0.1089 & 0.0749 & 0.0819 \\
        & w/o Interleaved   & 0.0609 & 0.0889 & 0.1099 & 0.0753 & 0.0821 \\
        & HoloRec & 0.0609 & 0.0894 & 0.1105 & 0.0761 & 0.0828 \\
        \bottomrule
    \end{tabular}
\end{table*}

\subsection{RQ1: Overall Performance}

Table~\ref{tab:main} presents the top-K recommendation performance of HoloRec and all baselines. Several observations can be made.
\begin{itemize}
    \item  HoloRec consistently outperforms every baseline across both datasets and all metrics, confirming the effectiveness of the proposed endogenous chain-of-thought mechanism. The improvement is particularly pronounced on the sparser Beauty dataset, where HoloRec achieves a 14.2\% relative gain in Hit@5 and 17.2\% in NDCG@5 over the strongest generative baseline TIGER. On the denser Instruments dataset, the gains are smaller but still positive, e.g., 4.4\% relative improvement in Hit@10. This pattern suggests that hierarchical semantic encoding and interleaved reasoning provide greater benefit when user interaction data is limited, as the coarse-to-fine structure offers a form of semantic regularization that mitigates data sparsity.
    
    \item Among generative recommendation baselines, codebook-based methods (TIGER and P5-SemID) consistently outperform the plain-text identifier method BIGRec. This aligns with prior findings that semantic IDs capture latent item relationships more effectively than raw text. HoloRec further improves upon TIGER by adding multi-granularity supervision and endogenous reasoning, indicating that flattening semantic IDs into a single granularity leaves room for hierarchical reasoning.
    \item Traditional sequential models such as SASRec and BERT4Rec perform reasonably on the denser Instruments dataset but fall behind on Beauty. Their lack of explicit semantic structure makes them more vulnerable to data sparsity, whereas HoloRec’s holistic encoding provides a more robust inductive bias.
\end{itemize}
 
Notably, the advantage of HoloRec is not merely a result of increased model capacity. As shown in the complexity analysis, the parameter overhead relative to TIGER is marginal, and the inference cost grows only by a constant factor. Therefore, the performance gains are primarily attributed to the hierarchical representation and the integrated reasoning mechanism rather than sheer model size. These results validate that explicit hierarchical semantics and endogenous chain-of-thought reasoning are effective directions for improving generative recommendation, especially in sparse scenarios.

\subsection{RQ2: Ablation Study}

\begin{figure}[tp]
\centering
\includegraphics[width=0.45\textwidth]{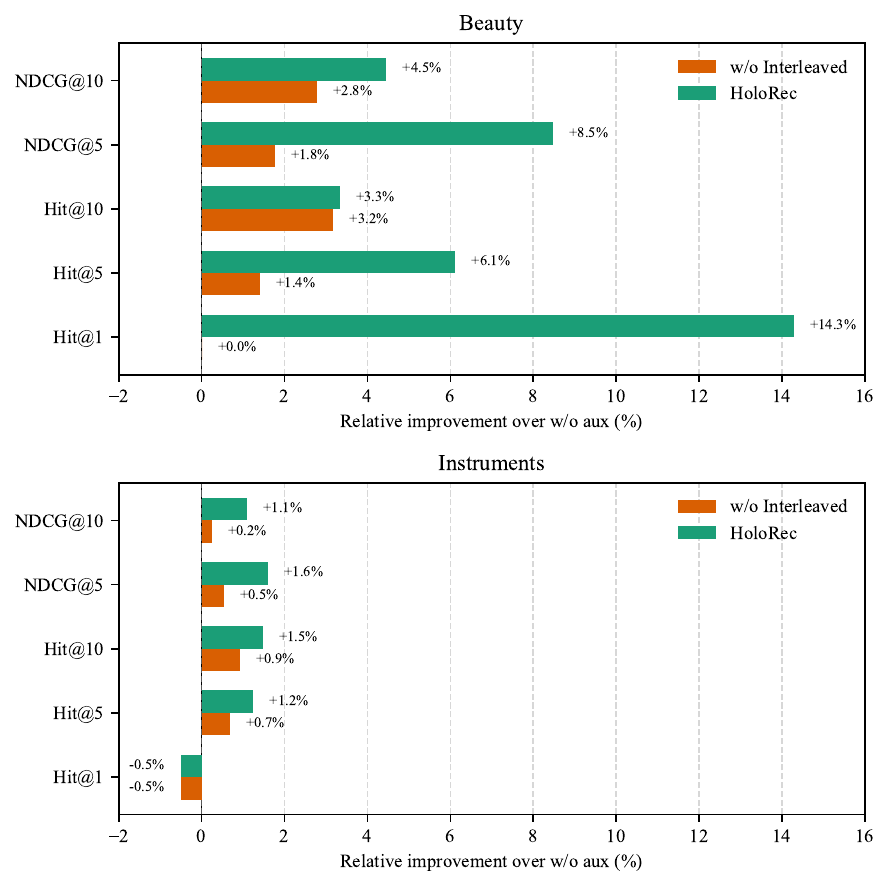}
\caption{Relative improvement (\%) over the baseline (w/o aux) on Beauty and Instruments. Orange: w/o Interleaved; blue‑green: HoloRec; red bars indicate negative changes. Both subplots share the same x‑axis range (–2\% to 16\%). Exact percentage labels are shown on each bar.}
\label{fig:improvement}
\end{figure}

We isolate the contribution of each core component in HoloRec by comparing three variants: \textit{w/o aux} uses only the finest granularity encoding without any coarse-grained supervision; \textit{w/o Interleaved} adds multi-granularity supervised alignment but removes the interleaved reasoning mechanism; and \textit{HoloRec} is the full model with both multi-granularity alignment and interleaved reasoning. Table~\ref{tab:method_comparison} reports the results. Figure~\ref{fig:improvement} visualizes the relative improvements over the baseline across both datasets. The key observations are as follows.

\begin{itemize}
    \item \textbf{Effect of multi-granularity supervision.} Comparing \textit{w/o Interleaved} with \textit{w/o aux} shows consistent but modest improvements, particularly on recall-oriented metrics such as Hit@10 rather than precision-oriented ones like Hit@1. This indicates that coarse-grained labels, which encode category-level information, help the model retrieve a broader set of plausible items without overly sharpening the top prediction.

    \item \textbf{Effect of interleaved reasoning.} Comparing \textit{HoloRec} with \textit{w/o Interleaved} reveals a different pattern. The gains are more pronounced for top-ranked recommendations, i.e., Hit@1 and Hit@5, while recall metrics remain nearly unchanged. This suggests that interleaved reasoning primarily refines the model’s ability to rank the first few positions by injecting coarse semantics step by step into the fine-grained decoding process, thereby imposing a coherent reasoning chain.

    \item \textbf{Modulation by data sparsity.} All improvements are substantially smaller on the denser Instruments dataset than on Beauty. This confirms that hierarchical reasoning provides stronger regularization and guidance when user interactions are scarce, whereas dense settings allow even a flat model to capture sufficient item relationships.
\end{itemize}

Both components are essential but play complementary roles: multi-granularity supervision broadens semantic awareness, while interleaved reasoning sharpens top recommendations. Their combined effect is most valuable in sparse scenarios.


\begin{figure}
    \centering
    \includegraphics[width=\linewidth]{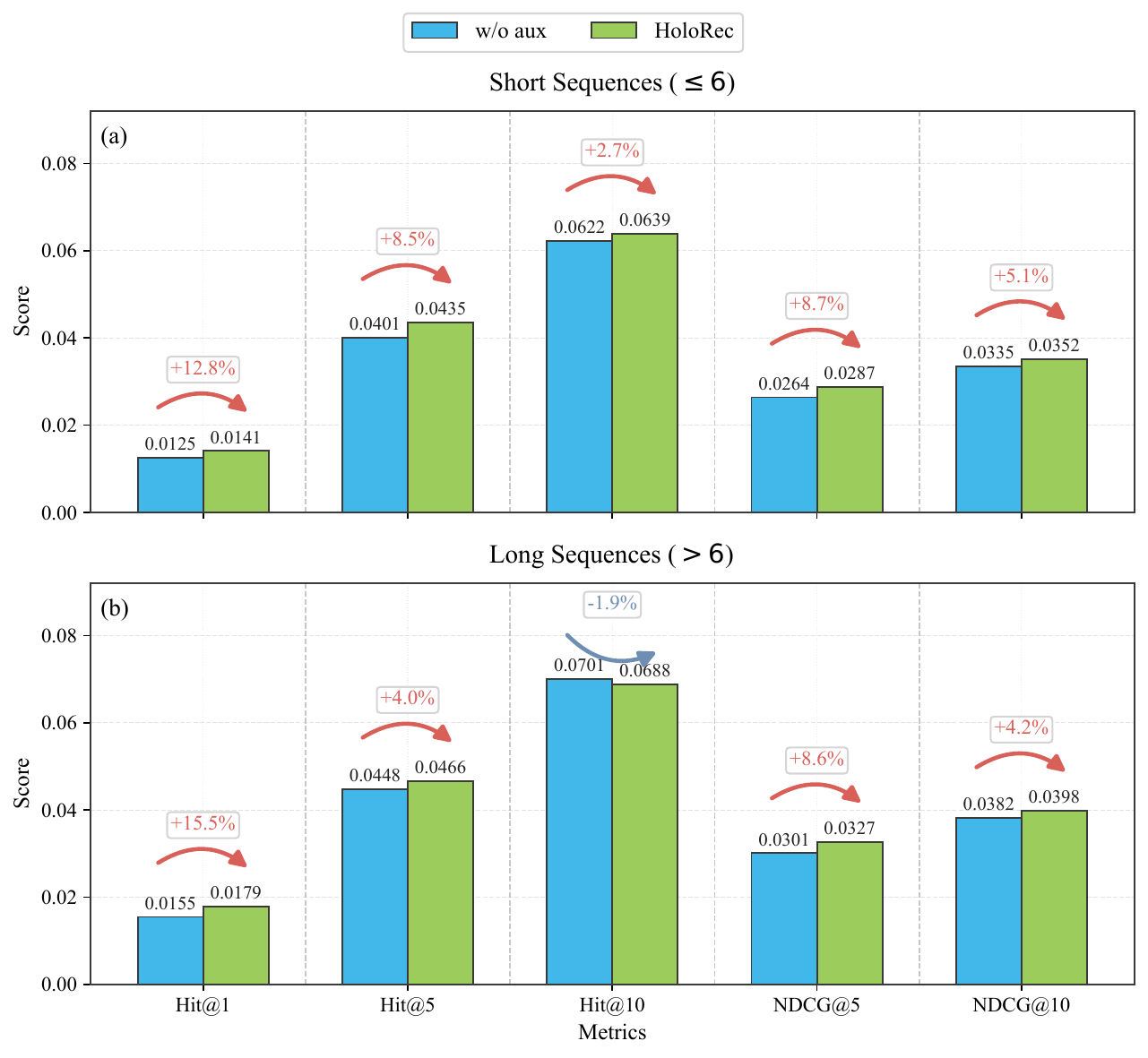}
    \caption{Performance comparison by sequence length on the Beauty dataset}
    \label{fig:beauty_length_fig}
\end{figure}

\subsection{RQ3: Analysis under Different Sequence Lengths}
We further analyze the impact of user history length on the Beauty dataset, splitting users into short sequences (\(\le 6\) interactions) and long sequences (\(> 6\) interactions). Table~\ref{tab:beauty_length} compares \textit{w/o aux} and HoloRec.

\begin{itemize}
    \item \textbf{Top-ranked recommendation gains.} HoloRec consistently improves Hit@1 for both short-sequence and long-sequence users. The relative gain is actually larger for long sequences, indicating that the interleaved reasoning mechanism benefits more from richer historical context, where stepwise coarse-to-fine reasoning can leverage longer user behavioral patterns.

    \item \textbf{Recommendation diversity.}  
    For 
    long-sequence users, Hit@10 shows a slight decrease under HoloRec compared to \textit{w/o aux}, whereas short-sequence users experience a modest increase. This suggests that coarse-grained guidance, while sharpening the top prediction, may inadvertently restrict the diversity of the full recommendation list when the user history is long. A possible explanation is that the model becomes overly reliant on the injected coarse semantics, narrowing its exploration of alternative fine-grained items.
\end{itemize}

Thus, HoloRec is particularly suitable for scenarios where top-1 accuracy is critical, e.g., limited-screen displays or voice assistants, but a hybrid or adaptive gating might be needed when diversity is paramount.

\begin{figure}
    \centering
    \includegraphics[width=\linewidth]{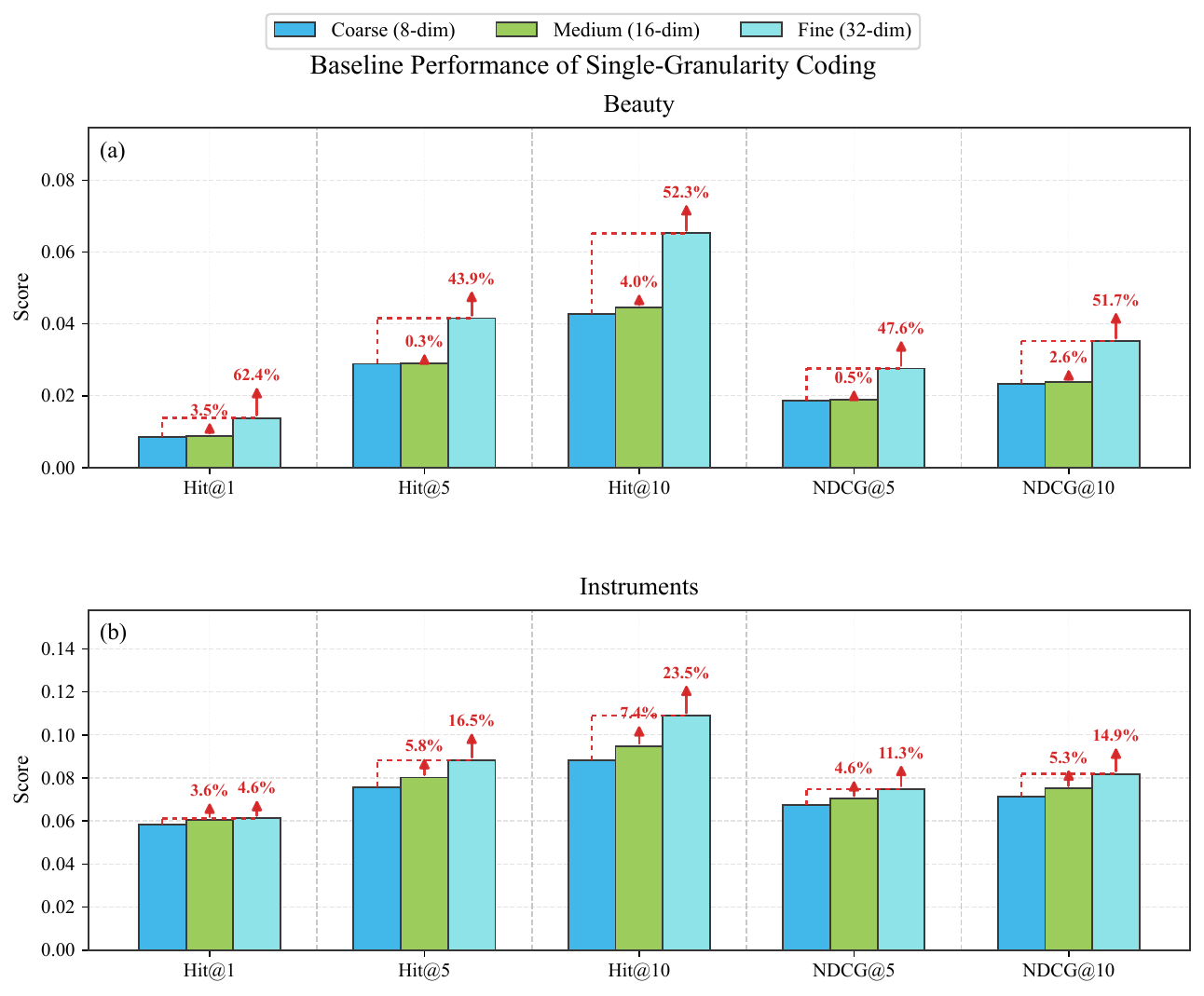}
    \caption{Baseline Performance of Single-Granularity Coding}
    \label{fig:granularity_comparison_fig}
\end{figure}

\subsection{RQ4: Analysis of Different Granularity Encodings}

We separately use the coarse granularity (8 dimensions), medium granularity (16 dimensions), and fine granularity (32 dimensions) encodings from the holistic semantic encoding to perform recommendation, without introducing joint multi-granularity supervision. The results are reported in Table \ref{tab:granularity_comparison}. On the Beauty dataset, the fine-grained encoding achieves a Hit@10 of 0.0652, substantially higher than the medium-grained 0.0445 and the coarse-grained 0.0428. On the Instruments dataset, the fine-grained encoding also outperforms the other two. This indicates that the fine-grained encoding preserves richer semantic information and is suitable as the target granularity for final recommendation output. Although the coarse-grained and medium-grained encodings have limited standalone performance, they serve effectively as auxiliary supervision signals to guide the model in learning hierarchical structures.


\begin{figure}
    \centering
    \includegraphics[width=\linewidth]{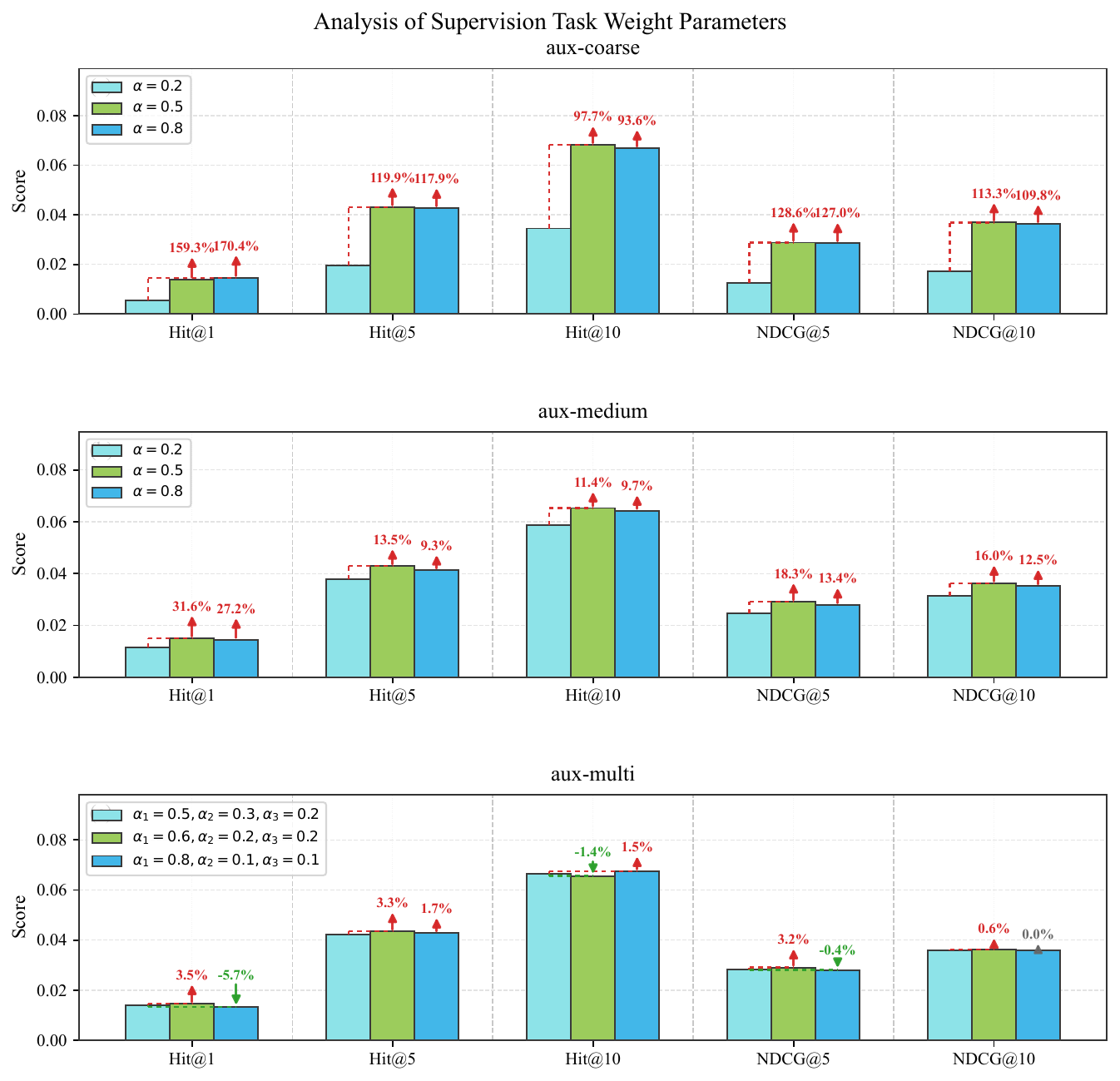}
    \caption{Analysis of Supervision Task Weight Parameters}
    \label{fig:alpha_analysis_fig}
\end{figure}

\begin{table*}[tp]
\centering
\caption{Comparison of Time and Space Complexity}
\label{tab:complexity}

\renewcommand{\arraystretch}{1.2}
\begin{tabular}{lccc}
\toprule
\multicolumn{4}{c}{\textbf{Time Complexity}} \\
\midrule
\textbf{Method} & \textbf{Offline SID Training} & \textbf{Offline GR Training} & \textbf{Online Inference} \\
\midrule
TIGER & $O(L K D)$ & $O(T^2 d_{\text{model}})$ & $O(T^2 d_{\text{model}})$ \\
\midrule

w/o aux & \multirow{3}{*}{$O(M L K D)$} & $O(T^2 d_{\text{model}})$ & $O(T^2 d_{\text{model}})$ \\
w/o Interleaved & & $O(MT^2 d_{\text{model}})$ & $O(T^2 d_{\text{model}})$ \\
HoloRec & & $O(MT^2 d_{\text{model}})$ & $O(MT^2 d_{\text{model}})$ \\
\midrule
\multicolumn{4}{c}{\textbf{Space Complexity}} \\
\midrule
\textbf{Method} & \textbf{Offline SID Training} & \textbf{Offline GR Training} & \textbf{Online Inference} \\
\midrule
TIGER & $O(K D + D^2)$ & $O(d_{\text{model}}^2 L_{\text{layers}})$ & $O(B T d_{\text{model}})$ \\
\midrule
w/o aux & \multirow{3}{*}{$O(K D + M D^2)$} & $O(d_{\text{model}}^2 L_{\text{layers}})$ & \multirow{3}{*}{$O(B T d_{\text{model}})$} \\
w/o Interleaved & & $+ O(d_{\text{model}} |V_{\text{coarse}}|)$ & \\
HoloRec & & $+ O(d_{\text{model}} |V_{\text{coarse}}| + d_{\text{model}}^2)$ & \\
\bottomrule
\end{tabular}
\end{table*}

\subsection{RQ5: Analysis of Supervision Task Weight}

In the multi-granularity supervised alignment training, the hyperparameter \(\alpha\) balances the weights between the fine-grained prediction loss and the coarse-grained supervision loss. We test different \(\alpha\) values and different auxiliary supervision choices (aux-coarse, aux-medium, and aux-multi) on the Beauty dataset, with results shown in Table \ref{tab:alpha_analysis}. The overall performance is optimal under each configuration when \(\alpha=0.5\). Performance drops significantly when \(\alpha=0.2\), indicating that an excessively large weight for the coarse-grained supervision task may interfere with the core recommendation objective. Using aux-medium (16-dim) yields better Hit@1 and NDCG@5, making it suitable for scenarios that prioritize top-ranked accuracy. Using aux-coarse (8-dim) yields better Hit@10 and NDCG@10, making it suitable for scenarios emphasizing recall. The multi-granularity fusion supervision (aux-multi) achieves balanced performance across various metrics and can adapt to diverse recommendation scenarios.

\subsection{Theoretical Complexity Analysis}

In the offline semantic ID training phase, the time complexity of holistic semantic encoding is \(O(M L K D)\), which is \(M\) times that of standard RQ-VAE. Since this phase is an offline one-time process, the overhead is acceptable. In the generative recommendation training and inference phases, the asymptotic time complexity of all methods is \(O(T^2 d_{\text{model}})\). The interleaved reasoning model adds one extra model forward pass per decoding step, roughly doubling the constant factor while preserving the theoretical order. In terms of parameter count, the holistic semantic encoding increases by approximately \(M D^2\) compared with RQ-VAE (with \(D=32\), about 3K additional parameters); in the generative recommendation model, the total increase from the coarse-grained prediction heads and the gating injection module accounts for less than 1\% of the backbone parameters. The inference memory footprint is nearly on par with the baseline, primarily \(O(B T d_{\text{model}})\). Taken together, HoloRec achieves improved recommendation accuracy while maintaining computational efficiency comparable to the baselines.

\section{Conclusion}
\label{sec:conclusion}
This paper addresses the challenges in generative recommendation systems where item semantic representations lack a hierarchical structure for reasoning, chain-of-thought data is scarce, and the reasoning process is disconnected from generation. We propose HoloRec, an endogenous chain-of-thought recommendation mechanism. Key contributions include a holistic semantic encoding method that constructs a hierarchical discrete encoding matrix, a non-thinking mode with multi-granularity supervised alignment, and an interleaved reasoning model that injects coarse-grained semantics into fine-grained decoding. Experiments on public datasets demonstrate that HoloRec consistently outperforms strong baselines, and ablation studies confirm the effectiveness of each component.

Future work includes extending holistic encoding to multimodal scenarios, designing more efficient reasoning mechanisms, and validating the method on industrial-scale datasets.

\subsection*{Acknowledgments}
This research was supported by the National Key Research and Development Program of China under Grant No. 2024YFC3307400, within the key special project for Social Governance and Smart Society Technology Support.

\bibliographystyle{ACM-Reference-Format}
\bibliography{sample-base}

\appendix

\end{document}